# The plasma β evolution through the solar corona during solar cycles 23 and 24

Jenny Marcela Rodríguez Gómez[1], Judith Palacios[2], Luis E. A. Vieira[3], and Alisson Dal Lago[3]

[1]Skolkovo Institute of Science and Technology (Skoltech), Space Center. The Autonomous structural subdivision,SD Moscow Nobel 3, Moscow
[2]Leibniz Institut für Sonnenphysik (KIS), Freiburg, Germany
[3]Instituto Nacional de Pesquisas Espaciais, INPE, Brasil

## ABSTRACT

The plasma β is important to investigate interchanging roles of plasma and magnetic pressure in the solar atmosphere. It can help to describe features over the photosphere and their changes at different heights. The goal is to obtain the plasma β variations through the solar corona during the solar cycles 23 and 24. The plasma β is reconstructed in different layers of the solar atmosphere. For this purpose, we use an updated version of COronal DEnsity and Temperature (CODET) model. In this version we selected different features in the solar atmosphere such as Quiet Sun (QS), faculae and Active Regions (ARs). We calculate the β variations at different layers in the solar corona (R= 1.14, 1.19, 1.23, 1.28, 1.34, 1.40, 1.46, 1.53, 1.61, 1.74, 1.79, 1.84 and 1.90 $R_\odot$). In the photosphere we use temperature values from FALC model to obtain plasma β in QS and faculae. Additionally, variations of the magnetic and kinetic pressure were modelled during the last solar cycles at coronal heights.

## 1. INTRODUCTION

The structure of the solar atmosphere is complex, due to interchanging roles of plasma and magnetic pressure. This behavior is usually described by the plasma β, because it relates the gas pressure (kinetic pressure) and magnetic pressure. The rapid decrease in gas pressure in the upper atmosphere suggests that $\beta \sim 1$ or smaller there. In layers between the photosphere and low chromosphere, β is roughly of order unity. The corona is a low-beta plasma. Then, the magnetic forces dominate over kinetic pressures. On the other hand, when β is high, kinetic forces dominate. In a low-beta plasma, the equilibrium field are force-free (Aschwanden 2005; Priest & Hood 1991). One of the most relevant papers on solar plasma β is Gary (2001), where a static magnetic field model of an active region was used for calculating the average and limits of the plasma β across the solar atmosphere. Recently, Wiegelmann et al. (2015) obtained some new constraints with IMaX data with a non-linear magnetostatic model for the photosphere and chromosphere on network elements. The plasma β may be an explanation for some solar processes, as penumbra formation (Bourdin 2017).

The plasma β is important to describe different physical phenomena in the solar atmosphere, such as solar wind and how is it expelled. In general, the wide range on β (sometimes of orders of magnitude in a given region) can be related to ejection events, reconnection processes and drive some dynamics in loops (Gary 2001). The dynamics of the β plasma in the solar atmosphere is not simple to describe, for example, in photosphere and lower chromosphere, low and high-β regions exist side by side along with non magnetic forces. The existence of the gradient of the plasma pressure and the gravity force are also considered as parameters in Wiegelmann et al. (2015).

Quiet Sun is an important feature in the solar atmosphere. It exhibits magnetic activity on a broad range of scales (Bellot Rubio & Orozco Suárez 2019). Faculae regions represent dispersed magnetic fields (van Driel-Gesztelyi & Green 2015). Their dynamics may contribute in some processes in solar chromosphere and corona. Then, the study of plasma β in these features can contribute to understand interchanging roles of magnetic and kinetic pressure at differents heights of solar atmosphere. In this paper we investigate the evolution of plasma β, magnetic and kinetic pressure over the last two solar cycles. Additionally, we describe changes in plasma β over different features: faculae, Quiet Sun and Active Regions and their evolution in times of maximum and minimum of the last solar cycles.

We structure the paper as follows. In Sect. 2, we describe the plasma β model. In Sect. 3, Plasma β evolution during the solar cycles 23 and 24 is presented. In sect. 4, Plasma β at coronal heights is presented. In sect. 5, variations in magnetic and kinetic pressure are described. Finally, we present the discussion and concluding remarks in Sect. 6.

## 2. THE PLASMA β MODEL

The plasma β is described through the magnetic and the kinetic pressure. In our model, we assume flux tubes as pressure balanced elements between the inner and outer boundaries. These tubes are assumed not to be curved and thus have no magnetic tension. Then, it is possible to describe the plasma β as:

$$\beta = \frac{2 N_e K_B T_e}{B_m^2 / 8\pi} \qquad (1)$$

Where $N_e$ and $T_e$ corresponds to the average electronic density and temperature in each layer through the solar corona. $B_m = \frac{B(r,\theta,\phi)}{B_s}$ is the magnetic field amount each pixel, $B_s = 4.4080\ G$ obtained in Rodríguez Gómez et al. (2018), $K_B = 1.38\ x\ 10^{-16}\ erg/K$ is the Boltzmann constant. In this approach, we could not directly employ the magnetic field strength density from MDI/SOHO (Scherrer et al. 1995) and HMI/SDO (Scherrer et al. 2012). These data are assimilated into the flux transport model of Schrijver (2001). The structure of the coronal magnetic field is estimated to employ a Potential Field Source Surface[1] (Schrijver 2001; Schrijver & De Rosa 2003). It allows us to obtain $B(r, \theta, \phi)$ values from $1\ R_\odot$ to $2.5\ R_\odot$.

---

[1] Solarsoftware. www.lmsal.com/solarsoft

The density and temperature description were presented in Rodríguez Gómez et al. (2018) on CODET model. For this work, we perform an update in CODET model. In the current version, the model is available to describe some features in the solar atmosphere, as faculae, Active regions (ARs) and Quiet Sun regions (QS). We evaluate the behavior of plasma β in these regions along with the kinetic and magnetic pressure during two solar cycles. For these purposes, we defined some conditions to select different features in the solar atmosphere (Table 1).

Table 1. Conditions to select different features in the solar atmosphere, through different magnetic field thresholds.

| Feature | Magnetic field regime (G) |
|---|---|
| Quiet Sun Regions (QS) | 5 − 50 |
| Faculae | 50 − 150 |
| Active Regions (ARs) | 500 − 4000 |

These conditions were applied in the photosphere ($1\,R_\odot$) and the model was applied to these different conditions (magnetic regime) for all layers. Figure 1 shows binary maps for Active Regions, faculae and Quiet Sun on Apr. 17 (2014). These conditions were included in the update of CODET model. We perform a spatial averaging at different heights due to structure of the Potential Field Source Surface (PFSS). We obtain a mean value of magnetic field, density and temperature at each height. Using these values we calculated the plasma β in each feature at each height in solar atmosphere for each day. For our goals, it would be useful to distinguish variations in plasma β during the solar cycle. Therefore we select some days during the maximum and minimum of solar cycles 23 and 24. These results will be presented in the following sections.

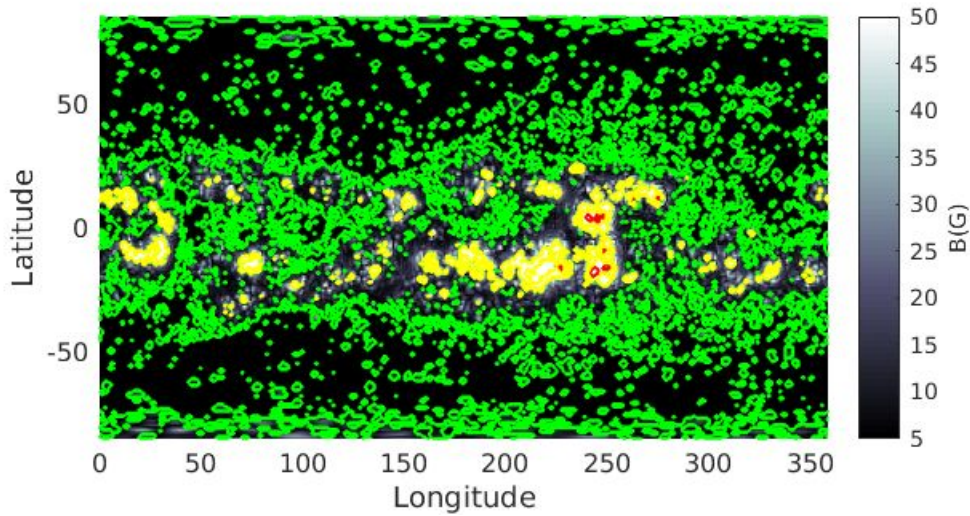

**Figure 1.** Selection conditions different features applied on Apr. 17 (2014). Magnetic field at the photosphere height. The conditions were applied to select Active regions (AR) (red line contours), Quiet Sun (green line contours) and Faculae regions (yellow line contours).

# 3. PLASMA β EVOLUTION DURING THE SOLAR CYCLES 23 AND 24

The plasma β was calculated from Jul. 01 (1996) to Aug. 23 (2018). The average plasma β value was calculated at coronal heights 1.14, 1.19, 1.23, 1.28, 1.34, 1.40, 1.46, 1.53, 1.61, 1.74, 1.79, 1.84 and 1.90 $R_\odot$, following Rodríguez Gómez et al. (2018) the performance of the CODET model is suitable in these layers of the solar corona.

Figure 2 shows the evolution of the plasma β over the last two solar cycles per day. The plasma β was computed for all regions during the two solar cycles. After that, it was sorted for QS, faculae and AR (Panels figure 2). In selected layers in the solar corona we obtain β > 1. These results are in agreement with expected values at coronal heights. The contributions from faculae and Quiet Sun regions are higher compared to Active Regions during both solar cycles. In general, plasma β and the contributions from faculae and Quiet Sun follow the solar cycle.

**Table 2.** Maximum and minimum values of plasma $\beta$.

| Feature | $\beta$ minimum value | Height | Date | $\beta$ maximum value | Height | Date |
|---|---|---|---|---|---|---|
| Total | 1.07 | 1.40 $R_\odot$ | Jun. 19 (2016) | 34.46 | 1.14 $R_\odot$ | Oct. 24 (2014) |
| Faculae | 0.31 | 1.40 $R_\odot$ | Jan. 27 (2009) | 9.20 | 1.14 $R_\odot$ | Oct. 24 (2014) |
| Active Regions | 0.12 | 1.14 $R_\odot$ | Oct. 28 (2014) | 13.71 | 1.14 $R_\odot$ | May. 29 (2001) |
| Quiet Sun | 1.08 | 1.40 $R_\odot$ | Sep. 23 (2013) | 9.28 | 1.14 $R_\odot$ | Oct. 24 (2014) |

Table 2 shows maximum and minimum values of plasma β without distinguishing any feature (total), faculae, Active Regions and Quiet Sun. Also, their respective heights and dates when the maximum and minimum values occurs. Minimum values appear at a height of 1.40 $R_\odot$ for total, faculae and Quiet Sun. However, for Active Regions the minimum values appear at 1.14 $R_\odot$. Maximum values occur at 1.14 $R_\odot$ in all cases. The maximum and minimum values for the entire atmosphere occurs in the solar cycle 24. While, Faculae the minimum value appears on the minimum of solar cycle and the maximum value occurs in the solar cycle 24. Active regions shows a minimum value on solar cycle 24 and the maximum value on solar cycle 23. This related to magnetic activity variations during each solar cycle. Quiet Sun regions shows maximum and minimum values during solar cycle 24.

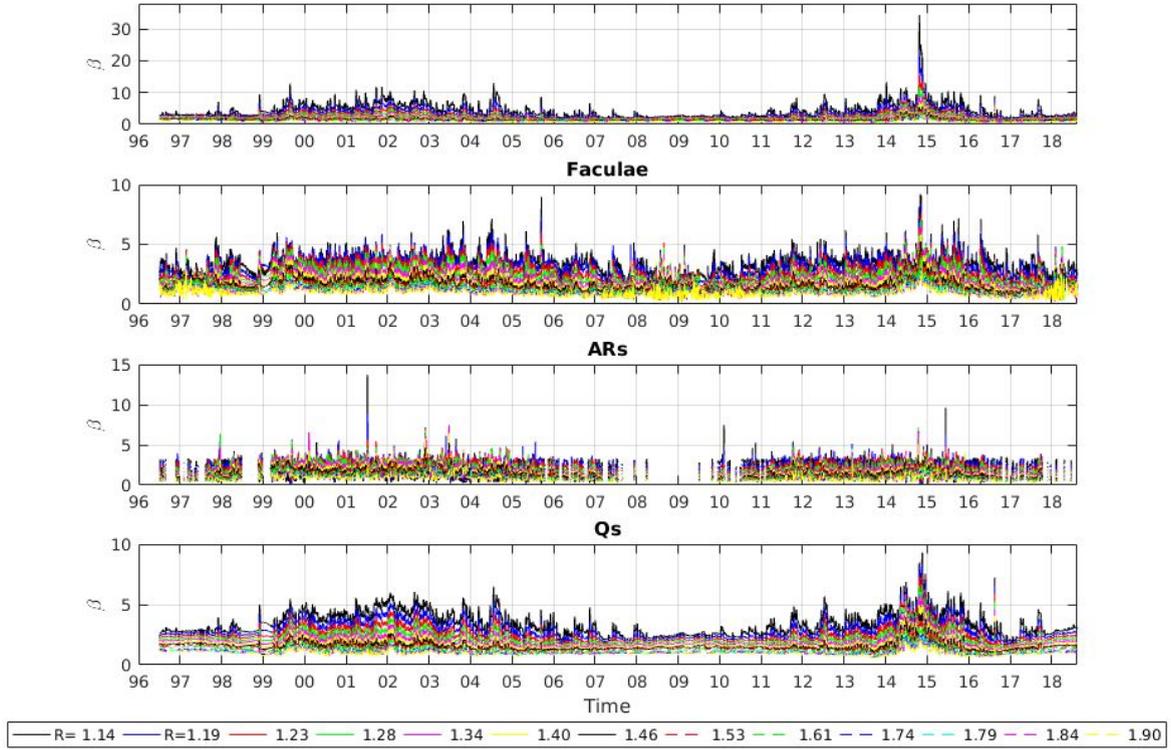

**Figure 2.** Plasma β variation through different layers R= 1.14, 1.19, 1.23, 1.28, 1.34, 1.40, 1.46, 1.53, 1.61, 1.74, 1.79, 1.84 and 1.90 $R_\odot$ in the solar corona, from Jul. 01 (1996) to Aug. 13 (2018). Top panel: plasma β evolution without distinguishing any feature. Middle panels: β over faculae regions and Active Regions (ARs). Bottom panel: plasma β over Quiet Sun regions (QS).

## 4. PLASMA β IN FUNCTION OF HEIGHT IN THE SOLAR ATMOSPHERE

To compute the plasma β in the photosphere, we use the temperature values from FALC model (Fontenla et al. 2011) for faculae and Quiet Sun at height of $1.0026\,R_\odot$. FALC model describes semi-empirical atmospheric models of solar features such as Quiet Sun and Faculae. These models are created assure a reasonable behavior of spectrally integrated radiative losses and to better match the SORCE/SIM observations (Fontenla et al. 2011). We take an average value and calculate the standard error. We were selected the models 1005 and 1008 for faculae and hot faculae. The average value of temperature for faculae correspond to $T_e = 2 x 10^5 \pm 2\,K$. For Quiet Sun, we selected the models 1000, 1001 and 1002 for Dark quiet sun inter-network, Quiet Sun inter-network and Quiet Sun network lane. The temperature average value for QS is $T_e = 2 x 10^5 \pm 427.8\,K$. Using this value of temperature from FALC model and the density profile related to magnetic field variations from CODET model (Rodríguez Gómez et al. 2018). We calculated the plasma β value for these features at different days during the maximum and minimum of solar cycles 23 and 24 (Table 3.).

**Table 3.** Plasma $\beta$ values in the solar photosphere for Quiet Sun and faculae regions.

| Date | $\beta$ Quiet Sun | $\beta$ Faculae |
|---|---|---|
| Nov. 21 (2001) | 1.05 | 1.41 |
| Apr. 17 (2014) | 1.04 | 1.38 |
| Dec. 29 (1996) | 0.78 | 1.38 |
| Dec. 22 (2008) | 0.74 | 1.43 |
| Jan. 08 (2009) | 0.73 | 1.02 |

Table 3 shows values of plasma β for Quiet Sun and faculae in selected days on solar cycles 23 and 24. However, modelling the ARs is more complex. We use some models to obtain temperature values, but they do not match to the expected values in the photosphere. Then, in this approach we do not use plasma β values over the ARs in the solar photosphere. In order to check the values of plasma β related to the different features obtained from our model, we decided to show the values corresponding to maxima and minima of solar cycles 23 and 24. Days were chosen randomly along the month from the 13-month averaged SSN series. Maximum of cycle 23: Nov. 21 (2001). Minimum of cycle 23: Dec. 29 (1996) and Dec. 22 (2008). Maximum of cycle 24: Apr. 17 (2014). Minimum of cycle 24: Jan. 8 (2009).

Values of β along maxima and minima of solar cycles 23 and 24 were plotted over the figure 3 of Gary (2001). The thick line represents the AR physical conditions, with magnetic field of 2500 G. The thin line represents a plage with 150 G.

This approach allows to describe correctly the plasma β at photospheric heights for Quiet Sun and faculae. These values correspond to the expected values from 0.2 to $10^2$ (Table 3). Figures 3 to 5 shows plasma β values at photospheric and coronal heights. Faculae, Quiet Sun and Active Regions were checked on selected days. Blue points shows plasma β on faculae regions, red points correspond to Quiet Sun regions and black points to Active Regions. We expected values of Active regions and faculae reach the left side, close to the tick line due to the large magnetic field involved. This is related to higher magnetic field regimes. On the right side close to the thin line, Quiet Sun values appear usually, due to corresponding small magnetic field values.

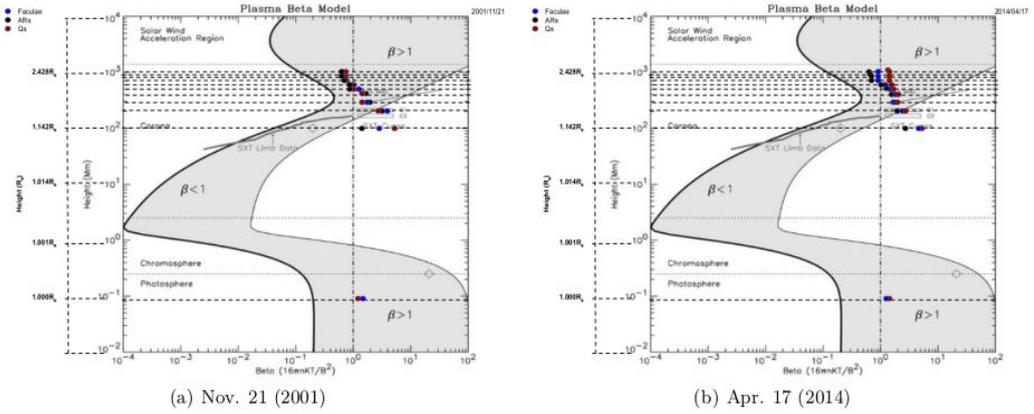

(a) Nov. 21 (2001)    (b) Apr. 17 (2014)

**Figure 3.** Plasma $\beta$ values in some days in the maxima cycles 23 and 24: Nov. 21 (2001) and Apr. 17 (2014). For different features: faculae (blue points), ARs (black points) and QS (red points).

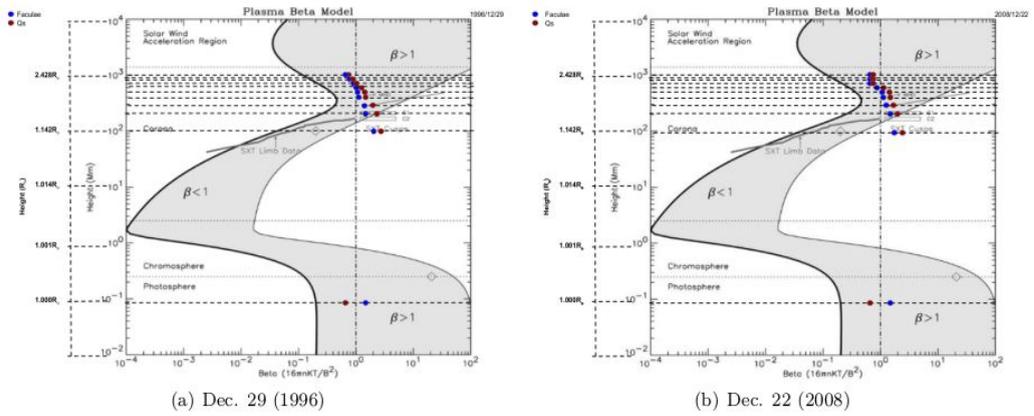

(a) Dec. 29 (1996)    (b) Dec. 22 (2008)

**Figure 4.** Plasma $\beta$ values in specific days of minimum of solar cycle 23: Dec. 29 (1996) and Dec. 22 (2008) for different features: faculae (blue points), and QS (red points).

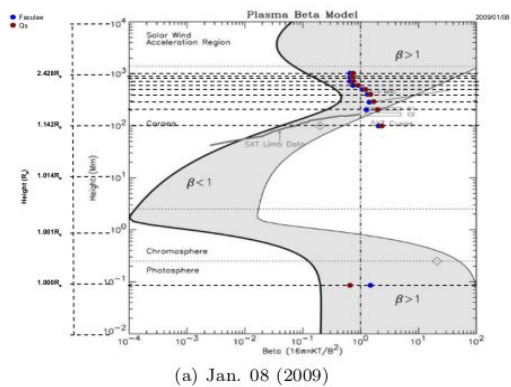

(a) Jan. 08 (2009)

**Figure 5.** Plasma $\beta$ values in Jan. 08 (2009) during the minimum of solar cycle 24, for different features: faculae (blue points) and QS (red points).

## 5. PRESSURE DESCRIPTION

The pressure values were examined during the last solar cycles. The main goal is to evaluate how the magnetic and kinetic pressure come into play during the each solar cycle. We define the magnetic pressure $P_m$:

$$P_m = B_m^2/8\pi \quad [dyn/cm^2] \qquad (2)$$

where $B_m$ is the magnetic field in each pixel and the kinetic pressure $P_k$ as:

$$P_k = 2N_e K_B T_e \quad [dyn/cm^2] \qquad (3)$$

where $N_e$ is the density and $T_e$ is the temperature average in each layer, from CODET model (updated version) for coronal heights.

Figure 6 shows magnetic pressure variations along solar cycles 23 and 24 at coronal heights. The magnetic pressure follows the solar cycle, showing higher values on solar cycle 23 compared to solar cycle 24. Higher contributions are due to faculae and Quiet Sun regions during booth cycles. Figure 7 shows the kinetic pressure along the solar cycles 23 and 24. The kinetic pressure follows the solar cycle. Higher contributions are due to faculae and Quiet Sun regions.

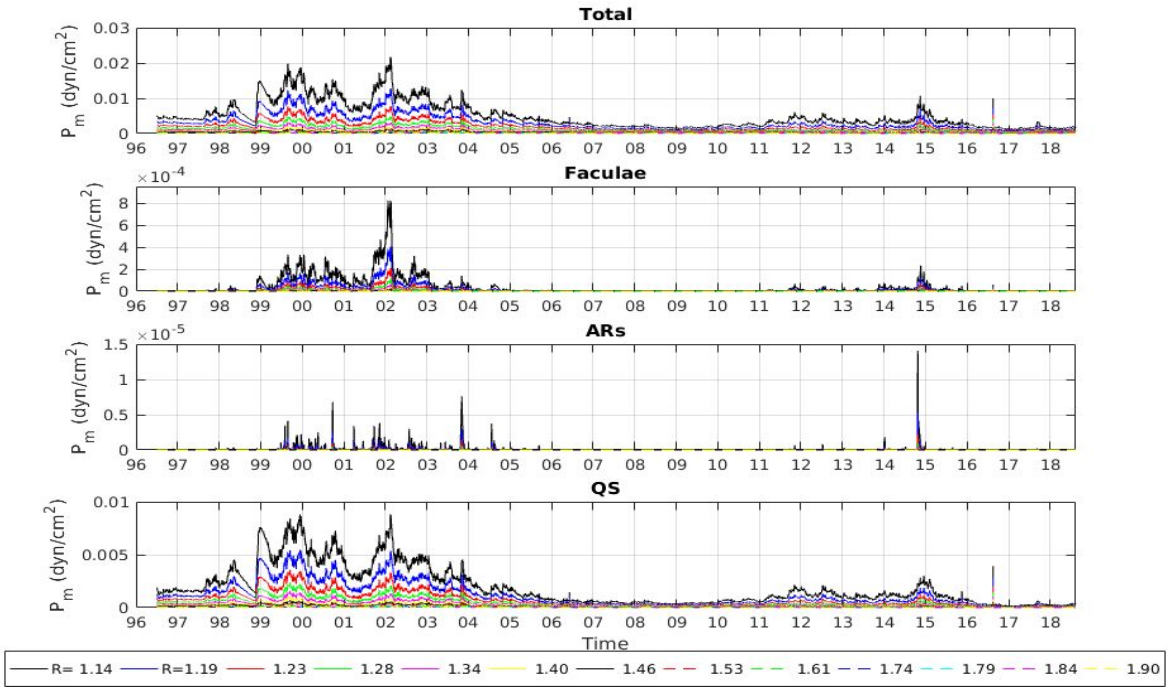

**Figure 6.** Magnetic pressure along the last two solar cycles, through different layers R= 1.14, 1.19, 1.23, 1.28, 1.34, 1.40, 1.46, 1.53, 1.61, 1.74, 1.79, 1.84 and 1.90 $R_\odot$ in the solar corona, from Jul.01 (1996) to Aug. 13 (2018). Top panel: total magnetic pressure. Middle panels: magnetic pressure on faculae and Active Regions regions. Bottom panel: magnetic pressure on Quiet Sun regions (QS).

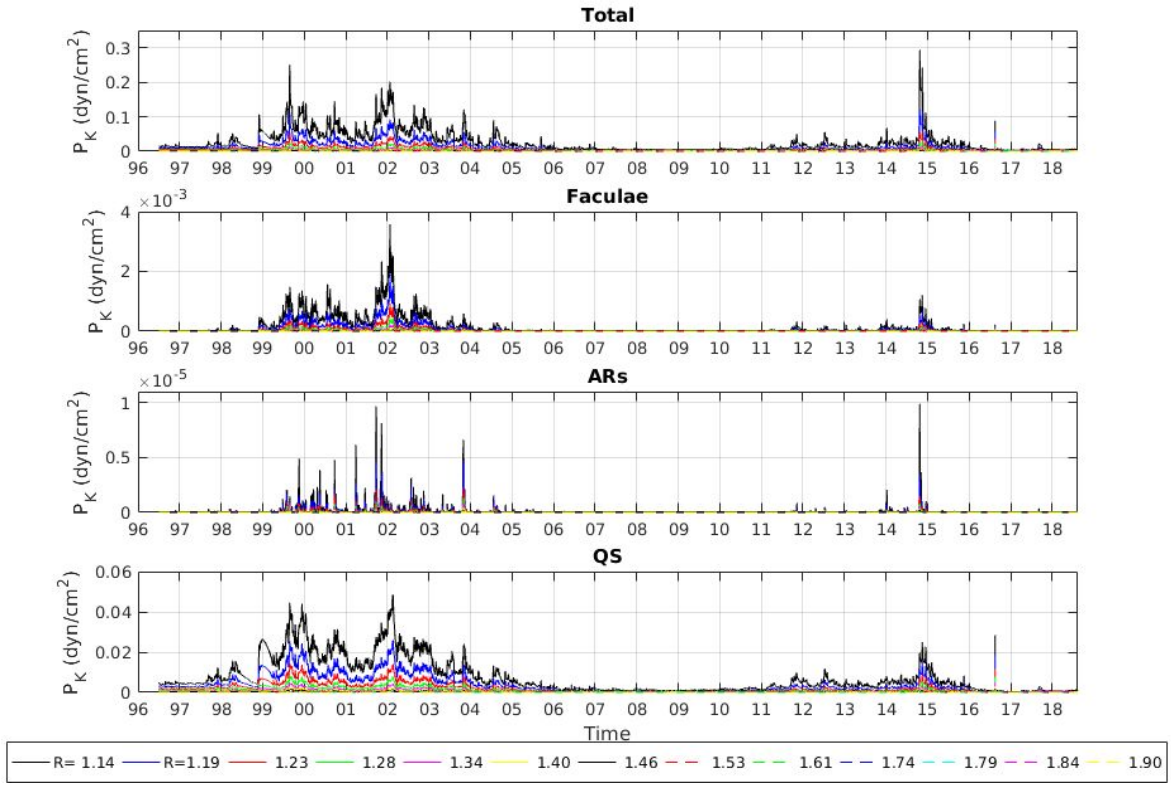

**Figure 7.** Kinetic pressure along the last two solar cycles. Through different layers R= 1.14, 1.19, 1.23, 1.28, 1.34, 1.40, 1.46, 1.53, 1.61, 1.74, 1.79, 1.84 and 1.90 $R_\odot$ in the solar corona, from Jul. 01 (1996) to Aug. 13 (2018). Top panel: total kinetic pressure. Middle panels: kinetic pressure on faculae and Active Regions and bottom panel: kinetic pressure on Quiet Sun regions (Qs).

## 6. DISCUSSION AND CONCLUDING REMARKS

The plasma β descriptions in different locations of the solar corona thus strongly depends on the magnetic field model. In our case, we use the Potential Field Source Surface (PFSS) to describe magnetic field at different heights of the solar corona. Besides, plasma β indeed depends on density and temperature in solar corona. These quantities are considered a great challenge because they are not estimated directly in long time scales. Therefore, it is necessary use models to obtain a description of them. This is the case of CODET model.

This model allows to describe temperature and density in the solar corona. Their description is based on magnetic field through solar atmosphere. However, some techniques can help to obtain density and temperature at different heights in the solar corona. e.g., the density can be obtained from high resolution data and using tomographic inversion. The tomographic technique and specifically the differential emission measure tomography (DEMT) provides a reconstruction of the 3D distribution of the coronal electron density and temperature in the inner corona (Vásquez 2016; Aschwanden 2011).

This approach will be explored to obtain plasma β estimations in a future paper. In this work, the evolution of plasma β during the last solar cycles were investigated. Also, we analyze the variations of magnetic and kinetic pressure in the last solar cycles. The global behavior was checked and the individual behavior in faculae, Quiet Sun and Active Regions. The plasma β description of the updated version of CODET model is in agreement within the limits and values presented in Gary (2001). Specifically, this version of CODET model is suitable to describe plasma β variations in the solar corona from $1.14\,R_\odot$ to $1.90\,R_\odot$ in different features as faculae, Quiet Sun (QS) and Active Regions (ARs).

We notice that an interesting behavior is shown during the last solar cycles, when we select different regions such as faculae, Active Regions and Quiet Sun. In Faculae and Quiet Sun regions, higher values of β in average are shown in solar cycle 24 compared to solar cycle 23. The solar cycle 24 is characterized by lower magnetic activity, but the presence of higher values in plasma β for faculae and Quiet Sun suggest that these features are predominant and they can drive some phenomena during this cycle. We suggest that the role of the gas pressure instead of strong magnetic forces is predominant in the solar cycle 24. In the case of Active Regions the plasma β follow the solar cycle evolution corresponding to higher values in the solar cycle 23 and lower values in the solar cycle 24 related to magnetic activity.

However, their values are lower compared to other features considered. It can be related to the selection conditions on magnetic field regimes (Table 1) that were sought in the photosphere ($1\,R_\odot$) after the flux model was applied. This means that these values are slightly lower than photospheric magnetograms.

We check some peak values showed on the different features (Figure 2). Plasma β on Quiet Sun, faculae and Active Regions shows some particular peaks which deviate from the average values. They are related to a lower temperature and density at coronal heights. A special case occurred on Aug. 16 (2016) on Quiet Sun feature. It is related to an issue from the magnetic flux transport model and it influences the temperature and density profiles. Plasma β in function of height in the solar atmosphere is explored on specific days. Two selected days during maximum of solar cycles 23 and 24. Nov. 21 (2001) and Apr. 17 (2014) were shown in Figure 3. Faculae and QS values in the photosphere are in accordance with Gary (2001). At $1.14\,R_\odot$ all regions in different magnetic intensity regimes, namely ARs, faculae and QS show values of plasma β outside of the limits proposed by Gary (2001). In another layers plasma β values are in agreement with Gary (2001). During selected days on solar maxima 23 and 24, we can follow the main Active regions with coronal height. For minimum activity, on the minimum of solar cycle 23 (Figure 4), we can describe only plasma β related to QS and faculae regions. We selected two days: Dec. 26 (1996) and Dec. 22 (2008). In photosphere and coronal heights values of plasma β are in agreement with Gary (2001). At $1.14\,R_\odot$ plasma β values are smaller than the proposed limits of Gary (2001). Bourdin (2017) in Fig 1, shows β in the upper boundary of the simulation ($10^2\,Mm$, which corresponds to $1.14\,R_\odot$). These values, for QS and AR, lies out of Gary's 150 G boundary, same as these values in this paper, in all the shown plots.

Due to the large possible range of β, dynamics and physical conditions, values can be different depending on observations and applied models (Bourdin 2017; Iwai et al. 2014). Also, we selected two days: Jul. 16 (2009) and Jan. 08 (2009), during the minimum of solar cycle 24. Plasma β were calculated for faculae and Quiet Sun regions. Their behavior is similar to minimum of solar cycle 23. In almost all explored days during times of solar minimum, QS shows higher values compared to faculae regions, presumably due to a greater presence of QS regions during times of solar minima.

The magnetic pressure (Figure 6) shows higher values during the solar cycle 23 compared to the solar cycle 24. It is related to the magnetic intensity of the solar cycle. The major contribution during the solar cycle 24 is related to QS and faculae regions. However, QS magnetic pressure is strong during this cycle. The kinetic pressure shows an interesting behavior in solar cycle 23 (Figure 7). Faculae and Quiet Sun regions have higher contribution into coronal dynamics. QS contribution is higher compared to faculae and ARs. The solar cycle 24 shows low activity intensity compared to solar cycle 23. However, the QS regions contribute in major part during this cycle. A pressure imbalance due to the higher value of kinetic pressure occurred and higher values of β appear. In order to maintain a positive total pressure, the background plasma pressure must be strong either the magnetic pressure is so weak that the average plasma β becomes very high.

These results give an interesting outlook about the solar cycle dynamics. The faculae and Quiet Sun regions drive variations in magnetic and kinetic pressure at coronal heights. However, it is important to know the behavior of plasma β at chromosphere and transition regions. Nevertheless the current description is not available for this purpose, but we expected to explore this on outcoming paper.

## ACKNOWLEDGMENT

J.R.G thanks the CNPq/Brazil project N 300596/2017 − 0. J. P. acknowledges the KIS 'Solare Atmosphäre Schwerpunkte, mit Magn.Aktiv.i.d.Chromosphäre + Korona'. We thank the anonymous reviewer their comments that helped improve this paper.## REFERENCES

Aschwanden, M. J. 2005, Physics of the Solar Corona. An Introduction with Problems and Solutions (2nd edition). 2011, Living Reviews in Solar Physics, 8, 5,doi: 10.12942/lrsp-2011-5

Bellot Rubio, L., & Orozco Suárez, D. 2019, Living Reviews in Solar Physics, 16, 1, doi: 10.1007/s41116-018-0017-1

Bourdin, P.-A. 2017, ApJL, 850, L29, doi: 10.3847/2041-8213/aa9988

Fontenla, J. M., Harder, J., Livingston, W., Snow, M., & Woods, T. 2011, Journal of Geophysical Research (Atmospheres), 116, D20108, doi: 10.1029/2011JD016032

Gary, G. A. 2001, solphys, 203, 71, doi: 10.1023/A:1012722021820

Iwai, K., Shibasaki, K., Nozawa, S., et al. 2014, Earth, Planets, and Space, 66, 149, doi: 10.1186/s40623-014-0149-z